\documentclass{article}
\usepackage{spconf} % ICASSP/IEEE style

\usepackage{amsmath,amsfonts}
\usepackage{graphicx}

\usepackage[table]{xcolor}
\usepackage{booktabs}
\usepackage{multirow}
\usepackage{makecell}
\usepackage{tabularx,array}

\usepackage[utf8]{inputenc} % keep for pdflatex
\usepackage[T1]{fontenc}
\usepackage{microtype}
\usepackage{textcomp}
\usepackage{enumitem}
\usepackage{siunitx}
\usepackage{xcolor}
\usepackage{url}

\newcolumntype{M}{>{\raggedright\arraybackslash}m{9mm}} % narrow column for model names
\newcolumntype{Y}{>{\centering\arraybackslash}X}

\newcommand{\Overlap}{overlap event}
\newcommand{\PreOverlap}{pre-\Overlap}
\newcommand{\PostOverlap}{post-\Overlap}

\usepackage[hidelinks]{hyperref}

\title{Full-Duplex-Bench v1.5: Evaluating Overlap Handling for Full-Duplex Speech Models}
\name{
\shortstack{Guan-Ting Lin$^1$, Shih-Yun Shan Kuan$^1$, Qirui Wang$^2$, Jiachen Lian$^3$, \\ Tingle Li$^3$, Shinji Watanabe$^4$, Hung-yi Lee$^{1,5}$}}
  
\address{\shortstack{$^1$National Taiwan University, $^2$University of Washington, $^3$UC Berkeley, $^4$Carnegie Mellon University,\\ $^5$NTU Artificial Intelligence Center of Research Excellence (NTU AI-CoRE)}}

\begin{document}
 
\ninept
\maketitle

\begin{abstract}
Full-duplex spoken dialogue systems promise to transform human-machine interaction from a rigid, turn-based protocol into a fluid, natural conversation. However, the central challenge to realizing this vision, managing \textbf{overlapping speech}, remains critically under-evaluated. We introduce \textsc{Full-Duplex-Bench v1.5}, the first fully automated benchmark designed to systematically probe how models behave during speech overlap. The benchmark simulates four representative overlap scenarios: user interruption, user backchannel, talking to others, and background speech. Our framework, compatible with open-source and commercial API-based models, provides a comprehensive suite of metrics analyzing categorical dialogue behaviors, stop and response latency, and prosodic adaptation. Benchmarking five state-of-the-art agents reveals two divergent strategies: a responsive approach prioritizing rapid response to user input, and a floor-holding approach that preserves conversational flow by filtering overlapping events. 
Our open-source framework enables practitioners to accelerate the development of robust full-duplex systems by providing the tools for reproducible evaluation\footnote{Code and data are available at \url{https://github.com/DanielLin94144/Full-Duplex-Bench}}.
\end{abstract}

\begin{keywords}
Full-Duplex Model, Spoken Dialogue System, Evaluation Benchmark
\end{keywords}

\section{Introduction}
\label{sec:intro}
Spoken dialogue systems are poised to evolve from command-and-response tools into genuine conversational partners. The cornerstone of this transition is \emph{full-duplex} capability—the ability to speak and listen simultaneously—mirroring the dynamics of human conversation~\cite{skantze2021turn}. This concurrent processing enables fluid interactions, such as mid-utterance corrections and backchannels, which are essential for applications demanding high responsiveness, from in-car assistants to real-time translation.

\begin{figure}[t]
  \centering
  \includegraphics[width=1\linewidth]{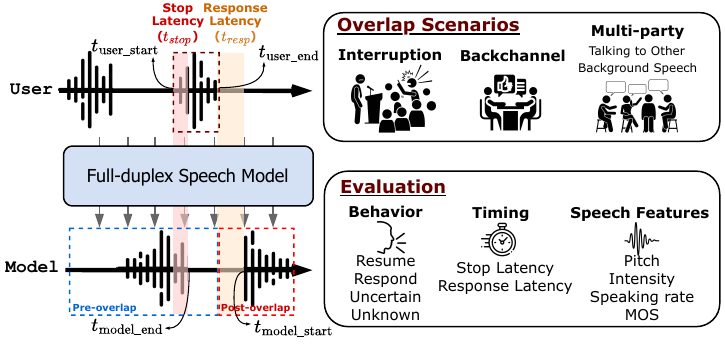}
  \vspace{-0.7cm}
    \caption{{\bf Overview of the evaluation framework for full-duplex speech models}. User speech (top) overlaps with model output (bottom) in four controlled scenarios. We analyze the model's post-overlap response across three dimensions: categorical behaviors, interaction timing, and adaptive speech features.}
    \vspace{-0.3cm}
  \label{fig:framework_1.5}
\end{figure}

The defining characteristic of natural dialogue is not the absence of interference, but the ability to manage overlapping speech in a meaningful way. Overlap is not an edge case but a common conversational event, accounting for over 40\% of turns~\cite{heldner2010pauses, schegloff2000overlapping, chime}. It plays diverse roles, from user interruptions and backchannels to side conversations and ambient speech~\cite{duncan1974signalling, sacks1974simplest}. A system that cannot handle overlap gracefully remains locked in a transactional, half-duplex paradigm, resulting in truncated responses, awkward silences, and degraded interaction quality.

Recent research has explored a variety of architectures to enable full-duplex behavior. Cascaded pipelines decompose the system into ASR, LLM, and TTS components and coordinate them with token-level control or time-sliced windows, as in FSM~\cite{fsm} and MiniCPM-Duplex/Duo~\cite{minicpm-duplex,minicpm-duo}. End-to-end approaches, such as dGSLM~\cite{dglsm}, SyncLLM~\cite{syncllm}, Moshi~\cite{moshi}, and NTPP~\cite{ntpp} learn to internalize turn-taking directly through joint speech modeling, while systems like SALMONN-omni~\cite{salmonn-omni}, and MinMo~\cite{minmo} introduce state tokens or dialogue managers to refine floor control. Despite this progress, reproducibility remains a challenge: only a few models (e.g., Freeze-Omni and Moshi) release public checkpoints, while most state-of-the-art systems—including Gemini Live, GPT-4o Realtime, and Nova Sonic—are accessible only through closed APIs.

Evaluation methodology has lagged even further behind. Mainstream speech benchmarks~\cite{superb,superb-prosody,air-bench,voxdialogue,sd-eval} focus on single-turn, half-duplex settings and fail to capture the dynamics of overlap phenomena. Human evaluations offer nuanced judgments but are expensive and difficult to reproduce. Corpus-level analyses rely on pause statistics and floor-transfer offset~\cite{dglsm}, which scale well but obscure scenario-specific behavior and ignore semantic appropriateness. Classifier-based approaches such as Talking Turns~\cite{talking-turns} automate turn-change detection but remain tied to specific training corpora, limiting their generality.

To address these gaps, we present \textsc{Full-Duplex-Bench v1.5}, an extension of our earlier work~\cite{full-duplex-bench}, which offers the first systematic, automated benchmark for overlap handling. Our framework streams audio in real time to both open-weight and API-based systems, introduces four controlled overlap scenarios—\textit{Interruption}, \textit{Backchannel}, \textit{Talking to Others}, and \textit{Background Speech}—and evaluates not only what a model says but also when and how it responds, through metrics that jointly measure dialogue behavior, timing, and prosodic adaptation. We benchmark five state-of-the-art systems and reveal two contrasting strategies for overlap management, quantifying their trade-offs and providing the community with a reproducible testbed for developing robust full-duplex dialogue systems.

\begin{figure}[t]
  \centering
  \includegraphics[width=1\linewidth]{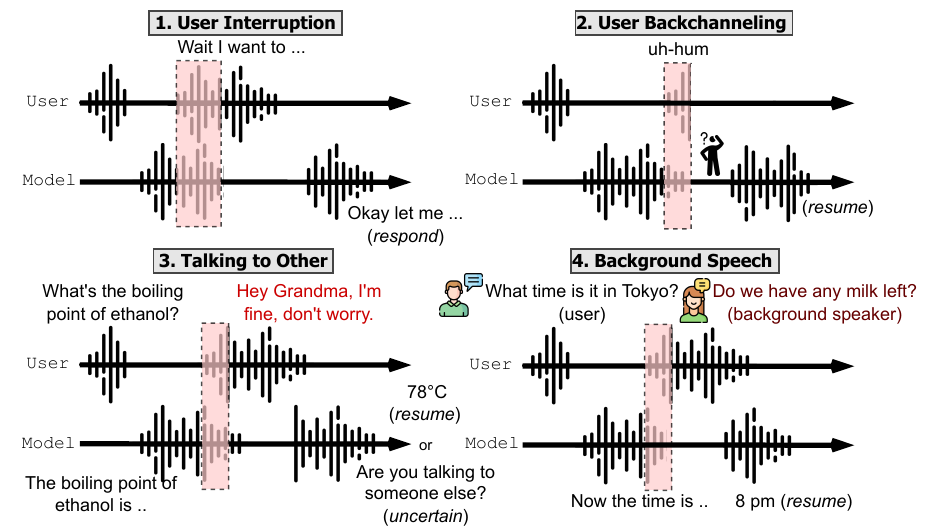}
  % \vspace{-0.7cm}
  \caption{{\bf Illustration of the four controlled overlap scenarios.}
User speech (top) and model speech (bottom) share a timeline:
(1) Interruption: user barges in with a new request;
(2) Backchannel: brief acknowledgment (uh-huh);
(3) Talking to Other: user addresses someone else;
(4) Background Speech: far-field third-party talk not meant for the model.}
  \label{fig:1.5_overview}
  % \vspace{-0.3cm}
\end{figure}

\section{\textsc{Full-Duplex-Bench-v1.5}}
\label{sec:benchmark}
% \textsc{Full-Duplex-Bench v1.5} conducts the evaluation of overlap handling through a standardized interaction framework, a set of controlled scenarios, and a suite of automated metrics.

\subsection{Interaction Framework}
Our framework simulates a live conversation by streaming pre-recorded user audio to a model while simultaneously capturing its output. This approach is model-agnostic, making no assumptions about internal architecture and thus supporting both open-source models and closed commercial APIs. The core evaluation primitive is a trial where a controlled \textit{\Overlap{}} is introduced while the model is speaking. The model's audio output is automatically segmented into \textit{\PreOverlap} and \textit{\PostOverlap} regions relative to the user's speech, enabling precise, localized analysis of its reaction (See the timeline illustration in Fig.\ref{fig:framework_1.5}). We adopt these names consistently throughout the paper; subsequent sections refer to them simply as the pre-overlap event and post-overlap event. The framework is fully extensible, allowing researchers to introduce new audio assets, languages, or custom scenarios.

% To naturally trigger the overlap event, we begin by using only the initial context turn and measuring each model's response latency and output speech duration to determine an appropriate timing for introducing the overlap speech. The results are presented in Table~\ref{tab:metrics_with_sd}. Based on the observed response latencies (1.18–2.25 seconds) and output durations (6.79–10.28 seconds), we insert a silence gap of approximately 4 seconds between the context and the onset of overlap speech. This timing ensures that the overlap reliably coincides with the model’s ongoing response, simulating realistic overlap scenarios.

% \begin{table}[h]
% \centering
% \caption{Average Output Length and Response Latency (mean ± std)}
% \label{tab:metrics_with_sd}
% \begin{tabular}{lcc}
% \toprule
% Model       & Output Length (s)      & Response Latency (s)     \\
% \midrule
% Freeze‑Omni & $9.57 \pm 1.93$      & $1.28 \pm 0.64$        \\
% GPT‑4o      & $9.18 \pm 2.40$      & $1.18 \pm 0.59$        \\
% Sonic       & $8.62 \pm 1.55$      & $2.25 \pm 0.58$        \\
% Gemini      & $6.79 \pm 2.47$      & $1.71 \pm 0.54$        \\
% Moshi       & $10.28 \pm 4.66 $      & $ 1.67 \pm 1.32 $        \\
% \bottomrule
% \end{tabular}
% \end{table}

\subsection{Controlled Overlap Scenarios}
\label{sec:benchmark-scenarios}
We design four scenarios, illustrated in Fig. \ref{fig:1.5_overview}, to probe distinct and essential conversational capabilities.

\subsubsection{User Interruption}
\label{sec:scenario-interruption}
\noindent\textbf{Capability Tested.}
Reactive turn-yielding and semantic repair. When a user barges in (e.g., ``Wait, I want to …"), an effective system must cede the floor rapidly and address the new query. This is critical in safety-conscious contexts like in-car navigation.

\noindent\textbf{Data.}
We synthesize 200 contextually relevant interruptions using the same speaker voice as the initial query, with no acoustic channel differences.

% ------------------------------------------------------------
\subsubsection{User Backchannel}
\label{sec:scenario-backchannel}
\noindent\textbf{Capability Tested.}
Filtering non-floor-taking cues. Listeners often produce short affirmations (e.g., ``uh-huh") to signal engagement, not to take the turn. The model must ignore these cues to maintain conversational flow.

\noindent\textbf{Data.}
We synthesize 99 backchannel utterances (e.g., \textit{yeah, right, mm-hmm}) from a curated list~\cite{jurafsky1997switchboard} using the same speaker voice.

% ------------------------------------------------------------
\subsubsection{User Talking to Others}
\label{sec:scenario-talkingtoother}
\noindent\textbf{Capability Tested.} 
Addressee detection and social appropriateness. When a user addresses another person (e.g., ``Hey Grandma, I'm fine. Don't worry."), the model must recognize that it is not the intended recipient and gracefully resume. This is crucial for multi-party settings.

\noindent\textbf{Data.}
We synthesize 100 utterances semantically directed to another person. To simulate the user speaking away from the device, we apply acoustic processing: volume reduced by 8 dB, a high-shelf filter attenuating frequencies above 4 kHz by 5 dB, and two reflections added at 45 ms (–6 dB) and 120 ms (–12 dB). This configuration simulates off-axis, far-field speech with early reflections, which is common in distant conversational setups~\cite{reverb}.

% ------------------------------------------------------------
\subsubsection{Background Speech}
\label{sec:scenario-background}
\noindent\textbf{Capability Tested.} 
Robustness to ambient acoustic interference. Real-world environments contain incidental speech not directed at the system. The model must remain focused on its task and not be derailed by this noise.

\noindent\textbf{Data.}
We synthesize 100 utterances with a different speaker voice on an unrelated topic. To simulate distant, environmental speech, we reduce the volume by 15 dB, apply a 3 kHz low-pass filter, and add echo with a 100 ms delay (–10 dB). We apply this spectral filtering and dynamic range compression to rigorously emulate distant, bandwidth-limited environmental speech~\cite{chime5}.

\subsection{Evaluation Metrics}

\subsubsection{Dialogue Behavior}
To understand a model's semantic strategy, we classify its \PostOverlap{} response into one of four categories using GPT-4o~\cite{gpt4}:
\begin{itemize}[topsep=0pt, noitemsep, leftmargin=*]
\item \textbf{Respond}: Addresses the content of the overlapping speech.
\item \textbf{Resume}: Ignores the overlap and continues its prior utterance.
\item \textbf{Uncertain}: Expresses confusion (e.g., “Could you repeat that?”).
\item \textbf{Unknown}: Produces an irrelevant response or remains silent.
\end{itemize}
Crucially, the text-based GPT-4o evaluator operates solely on ASR transcripts from (Parakeet-TDT) \cite{rekesh2023fast} to perform objective semantic categorization, strictly separating the evaluation modality (text) from the generation modality (audio) to minimize bias.

\subsubsection{Interaction Timing}
For each overlap event, we compute two timing measures (Fig. \ref{fig:framework_1.5} illustrates these time intervals):

\noindent\textbf{Stop latency} is the interval from the onset of overlapping user speech to the moment the model stops speaking:
\begin{equation}
t_{\text{stop}} = t_{\text{model\_stop}} - t_{\text{user\_start}} .
\label{eq:stop-latency}
\end{equation}
\noindent\textbf{Response latency} is the interval from the end of the overlapping speech to the model’s next utterance:
\begin{equation}
t_{\text{resp}} = t_{\text{model\_start}} - t_{\text{user\_end}} .
\label{eq:resp-latency}
\end{equation}
Here, $t_{\text{user\_start}}$ and $t_{\text{user\_end}}$ mark the start and end of user speech, while $t_{\text{model\_stop}}$ and $t_{\text{model\_start}}$ mark when the model stops and resumes speaking. Both $t_{\text{stop}}$ and $t_{\text{resp}}$ are in seconds, with speech activity detected by Silero-VAD~\cite{Silero-VAD}.
% 

% \subsubsection{Scenario-wise Expected Outcome}

% To clearly specify the behavioral targets for each overlap setting, 
% we summarize the expected post-overlap response, together with the 
% desirable stop and response latency profiles, in Table~\ref{tab:expected_outcome}. Stop latency reflects how readily the system yields the floor, whereas response latency measures how quickly it resumes after overlap. 
% These targets provide a reference against which the metrics in Table~\ref{tab:content_alltasks_with_latency_split} can be interpreted.

\begin{table}[t]
\centering
\scriptsize
\caption{Scenario-wise expected outcomes. Each row specifies the desired categorical behavior and latency for that overlap scenario.}
\label{tab:expected_outcome}
\begin{tabular}{lccc}
\toprule
\textbf{Scenario} & \textbf{Expected Behavior} & \textbf{Stop Latency} & \textbf{Response Latency} \\
\midrule
User Interruption & Respond & Low & Low \\
User Backchannel & Resume & High & Low \\
Talking to Others & Resume & High & Low \\
Background Speech & Resume & High & Low \\
\bottomrule
\end{tabular}
\end{table}

\subsubsection{Scenario-wise Expected Outcome}
We summarize the expected post-overlap behaviors and the corresponding desirable stop and response latency profiles in Table~\ref{tab:expected_outcome}.

In particular, \textit{Stop Latency} serves as an indicator of overlap awareness, with its optimal value varying by scenario: it should be \emph{short} when rapid yielding is required and may be \emph{longer} when the system is expected to hold the floor. In contrast, \textit{Response Latency} captures the post-overlap gap and is ideally minimized across all scenarios.

\noindent\textbf{User Interruption.} Expected behavior is \emph{Respond}: the system should promptly yield and address the overlapped intent. Thus we target a high \textit{Respond} rate with \textit{low Stop Latency} and \textit{low Response Latency}; continuing the prior turn or failing to react are primary errors.

\noindent\textbf{User Backchannel.} Expected behavior is \emph{Resume}: In conversation analysis, generic backchannels (e.g., “uh-huh”, “yeah”) conventionally invite the speaker to continue rather than take the floor~\cite{knudsen2020forgotten}. 
We target a high \textit{Resume} rate and minimal responses to backchannels; \textit{Stop Latency} should be as \emph{high} as possible (the model does not readily halt), while \textit{Response Latency} should remain \emph{low} to avoid unnecessary gaps.

\noindent\textbf{User Talking to Others \& Background Speech.} Expected behavior is \emph{Resume}, with brief addressee checks (\emph{Uncertain}) acceptable. We target high \textit{Resume} (with some \textit{Uncertain}) and near-zero responses to others; \textit{Stop Latency} should \emph{high} (holding the floor), whereas \textit{Response Latency} should be \emph{low}.

% \noindent\textbf{Background Speech.} Expected behavior is \emph{Resume}, optionally with brief verification (\emph{Uncertain}); the system should not respond to incidental speech. We target high \textit{Resume} (with some \textit{Uncertain}) and minimal responses to background; \textit{Stop Latency} should be \emph{high}, while \textit{Response Latency} should be \emph{low}.

%\section{Experimental Setup}
\section{Evaluation Setup}
We evaluate full‑duplex speech models via their streaming interfaces: 
\begin{itemize}[topsep=0pt, noitemsep, leftmargin=*]
\item \textbf{Freeze-Omni}~\cite{freeze-omni}: An open-source cascaded system with a frozen LLM, chunk-wise streaming speech input, and a classification head manages turn-taking. We deploy the official demo server locally.

\item \textbf{Moshi}~\cite{moshi}: An open-source, real-time speech-to-speech model with ``Inner Monologue'' for fluency and a multi-stream design for overlap. We use the official demo server.

\item \textbf{Gemini}~\cite{comanici2025gemini}: Google's commercial API, accessed via the gemini-2.0-flash-live-001 endpoint\footnote{\scriptsize \url{https://ai.google.dev/api/live}} with the Puck voice.

\item \textbf{Nova Sonic}~\cite{Intelligence2025}: Amazon's commercial service on AWS Bedrock, accessed via the amazon.nova-sonic-v1:0 endpoint\footnote{\scriptsize \url{https://github.com/aws-samples/amazon-nova-samples/tree/main/speech-to-speech/sample-codes/console-python}}.

\item \textbf{GPT-4o Realtime}~\cite{hurst2024gpt}: OpenAI's commercial API, accessed via the gpt-4o-realtime-preview-2024-12-17\footnote{\scriptsize \url{https://github.com/openai/openai-realtime-console}} with the alloy voice.
\end{itemize}

\begin{table}[t]
\centering
\scriptsize
\caption{Behavioral response distribution across overlap scenarios, with average stop and response latencies (s). Boldface indicates the best performance. Rows with a light-colored background highlight the \emph{desired behavior} for each scenario.}
\label{tab:content_alltasks_with_latency_split}
\begin{tabular}{llccccc}
\toprule
Scenario & Class / Metric & Freeze-Omni & Moshi & Gemini & Sonic & GPT-4o \\
\midrule
\multirow{6}{*}{\textsc{user\_intr}}
 & \cellcolor{blue!8}\textsc{Respond}$\uparrow$     & \cellcolor{blue!8}0.72 & \cellcolor{blue!8}0.50 & \cellcolor{blue!8}0.33 & \cellcolor{blue!8}0.24 & \cellcolor{blue!8}\textbf{0.78} \\
 & \textsc{Resume}$\downarrow$    
   & 0.12 
   & 0.26 
   & 0.55 
   & 0.71 
   & \textbf{0.10} \\
 & \textsc{Uncertain}$\downarrow$ & 0.03 & \textbf{0.00} & 0.01 & 0.01 & 0.02 \\
 & \textsc{Unknown}$\downarrow$   & 0.13 & 0.25 & 0.10 & \textbf{0.04} & 0.12 \\
\cmidrule(lr){2-7}
 & \textsc{Stop (s)}$\downarrow$  & 1.42 & 1.16 & 2.20 & 2.25 & \textbf{0.23} \\
 & \textsc{Resp (s)}$\downarrow$  & \textbf{1.35} & 1.47 & 2.62 & 2.75 & 1.50 \\
\midrule
\multirow{6}{*}{\textsc{user\_backch}}
 & \textsc{Respond}$\downarrow$   & 0.07 & 0.02 & 0.01 & \textbf{0.00} & 0.03 \\
 & \cellcolor{blue!8}\textsc{Resume}$\uparrow$      
   & \cellcolor{blue!8}0.80 
   & \cellcolor{blue!8}0.06 
   & \cellcolor{blue!8}0.93 
   & \cellcolor{blue!8}\textbf{0.98} 
   & \cellcolor{blue!8}0.70 \\
 & \textsc{Uncertain}$\downarrow$ & \textbf{0.02} & 0.00 & \textbf{0.02} & 0.00 & 0.01 \\
 & \textsc{Unknown}$\downarrow$   & 0.11 & 0.92 & 0.04 & \textbf{0.02} & 0.25 \\
\cmidrule(lr){2-7}
 & \textsc{Stop (s)}$\uparrow$  & \textbf{0.66} & 0.42 & \textbf{0.66} & 0.64 & 0.21 \\
 & \textsc{Resp (s)}$\downarrow$  & 2.16 & 3.00 & 2.45 & 1.45 & \textbf{1.32} \\
\midrule
\multirow{6}{*}{\textsc{talking\_other}}
 & \textsc{Respond}$\downarrow$   & 0.58 & 0.20 & \textbf{0.00} & 0.10 & 0.91 \\
 & \cellcolor{blue!8}\textsc{Resume}$\uparrow$      
   & \cellcolor{blue!8}0.25 
   & \cellcolor{blue!8}0.19 
   & \cellcolor{blue!8}\textbf{0.99} 
   & \cellcolor{blue!8}0.90 
   & \cellcolor{blue!8}0.02 \\
 & \textsc{Uncertain}$\uparrow$   & 0.00 & \textbf{0.02} & 0.00 & 0.00 & 0.01 \\
 & \textsc{Unknown}$\downarrow$   & 0.15 & 0.59 & 0.01 & \textbf{0.00} & 0.06 \\
\cmidrule(lr){2-7}
 & \textsc{Stop (s)}$\uparrow$  & 1.39 & 0.87 & 1.69 & \textbf{1.77} & 0.18 \\
 & \textsc{Resp (s)}$\downarrow$  & 1.32 & 2.38 & 1.78 & 2.04 & \textbf{1.16} \\
\midrule
\multirow{6}{*}{\textsc{bkg\_speech}}
 & \textsc{Respond}$\downarrow$   & 0.63 & 0.21 & 0.70 & \textbf{0.01} & 0.93 \\
 & \cellcolor{blue!8}\textsc{Resume}$\uparrow$      
   & \cellcolor{blue!8}0.25 
   & \cellcolor{blue!8}0.07 
   & \cellcolor{blue!8}0.30 
   & \cellcolor{blue!8}\textbf{0.98} 
   & \cellcolor{blue!8}0.04 \\
 & \textsc{Uncertain}$\uparrow$   & \textbf{0.01} & \textbf{0.01} & 0.00 & 0.00 & 0.00 \\
 & \textsc{Unknown}$\downarrow$   & 0.11 & 0.71 & \textbf{0.00} & 0.01 & 0.03 \\
\cmidrule(lr){2-7}
 & \textsc{Stop (s)}$\uparrow$  & 0.98 & 0.54 & 0.95 & \textbf{1.05} & 0.18 \\
 & \textsc{Resp (s)}$\downarrow$  & 1.60 & 1.62 & 2.38 & 2.76 & \textbf{1.26} \\
\bottomrule
\end{tabular}
\end{table}

\begin{table*}[t]
\centering
\scriptsize
\caption{Prosodic feature shifts (Pre $\rightarrow$ Post) during \textsc{Respond} trials for \textsc{user\_intr}. 
Arrows next to feature names indicate the \emph{expected} trend. Cells show actual change ($\Delta$) with $p$-values; significant increases are in {\color{green!50!black}green}, decreases in {\color{red!70!black}red}, and non-significant results are gray.}
\label{tab:intr_prepost}
\begin{tabular}{lccccc}
\toprule
Feature & Freeze-Omni & Moshi & Gemini & Sonic & GPT-4o \\
\midrule
\textbf{WPM $\uparrow$} & 
{\color{gray}+1.03} (.79) &
{\color{green!50!black}+59.55} (<.001) &
{\color{green!50!black}+18.09} (.037) &
{\color{gray}-11.63} (.14) &
{\color{green!50!black}+19.04} (<.001) \\
\textbf{Pitch Mean $\uparrow$} &
{\color{gray}-1.02} (.55) &
{\color{red!70!black}-8.94} (<.001) &
{\color{gray}+3.28} (.24) &
{\color{gray}+0.99} (.69) &
{\color{green!50!black}+5.60} (.002) \\
\textbf{Pitch SD $\uparrow$} &
{\color{gray}-0.08} (.93) &
{\color{red!70!black}-9.54} (<.001) &
{\color{gray}-0.78} (.73) &
{\color{gray}+0.57} (.71) &
{\color{green!50!black}+2.79} (.024) \\
\textbf{Intensity Mean $\downarrow$} &
{\color{red!70!black}-1.50} (.001) &
{\color{red!70!black}-4.02} (<.001) &
{\color{gray}-0.40} (.29) &
{\color{red!70!black}-1.88} (.010) &
{\color{gray}-0.07} (.75) \\
\textbf{Intensity SD $\uparrow$} &
{\color{green!50!black}+6.06} (<.001) &
{\color{green!50!black}+4.81} (<.001) &
{\color{gray}+0.07} (.80) &
{\color{green!50!black}+8.62} (<.001) &
{\color{red!70!black}-2.26} (<.001) \\
\textbf{UTMOSv2 $\rightarrow$} &
{\color{gray}-0.07} (.051) &
{\color{gray}-0.10} (.12) &
{\color{gray}-0.08} (.13) &
{\color{red!70!black}-0.21} (<.001) &
{\color{gray}+0.06} (.09) \\
\bottomrule
\end{tabular}
\end{table*}

\section{Results}
\label{sec:results}

\subsection{Scenario-wise Outcomes}
Table~\ref{tab:content_alltasks_with_latency_split} reports behavioral distributions and latencies across four overlap scenarios. Rows with shaded backgrounds indicate the desired outcome for each case.

\subsubsection{User Interruption}
The goal is rapid yielding and immediate handling of the new intent (\textsc{Respond} with low $t_{\text{stop}}$/$t_{\text{resp}}$). GPT-4o exhibits the strongest responsiveness (\textsc{Respond}=0.78) and extremely fast yielding ($t_{\text{stop}}{=}0.23$\,s), with Freeze-Omni closely following (\textsc{Respond}=0.72) and achieving the shortest $t_{\text{resp}}$ (1.35\,s). In contrast, Gemini and Sonic frequently continue their prior turn (\textsc{Resume}=0.55/0.71) and are slow to yield ($t_{\text{stop}}>2$\,s), reflecting excessive floor-holding under true interruption. Moshi is intermediate in timing but leaves many user intents unaddressed (\textsc{Respond}=0.50).

\subsubsection{User Backchannel}
Systems should continue speaking (\textsc{Resume}$\uparrow$), ignoring brief acknowledgments. Sonic and Gemini are most reliable (\textsc{Resume}=0.98 and 0.93, \textsc{Respond} close to 0), with Sonic also keeping response latency modest (1.45 s). Freeze-Omni maintains good continuity (\textsc{Resume}=0.80) but resumes more slowly ($t_{\text{resp}}=2.16$\,s). GPT-4o is highly sensitive ($t_{\text{stop}}=0.21$ s), suggesting over-eager halting even for short cues. Moshi is unstable (\textsc{Unknown}=0.92), failing to maintain turn continuity.

\subsubsection{User Talking to Others}
The preferred behavior is to hold the floor (\textsc{Resume}$\uparrow$) and avoid replying. Gemini performs nearly perfectly (\textsc{Resume}=0.99) with conservative yielding ($t_{\text{stop}}>1.7$\,s). Sonic is also strong (\textsc{Resume}=0.90). In contrast, GPT-4o treats most non-addressee speech as new intent (\textsc{Respond}=0.91), yielding almost immediately ($t_{\text{stop}}=0.18$\,s). Freeze-Omni shows mixed control (\textsc{Resume}=0.25), and Moshi again suffers from high uncertainty (\textsc{Unknown}=0.59).

\subsubsection{Background Speech}
The system should resume speaking quickly after noise but avoid treating it as an interruption. Sonic best matches this target (\textsc{Resume}=0.98, \textsc{Respond}=0.01), though with conservative gap management ($t_{\text{resp}}=2.76$\,s). In contrast, Gemini and Freeze-Omni often respond inappropriately (0.70/0.63), while GPT-4o nearly always yields (\textsc{Respond}=0.93) with near-zero stop latency, again over-accommodating. Moshi remains unstable (\textsc{Unknown}=0.71).

\subsection{Cross-Model Patterns and Error Modes}
A consistent trade-off emerges between rapid responsiveness and robust floor control. GPT-4o excels at true interruptions (fastest $t_{\text{stop}}$ and highest \textsc{Respond}), but systematically misfires under \textsc{talking\_other} and \textsc{bkg\_speech}. Sonic and Gemini exhibit strong addressee discrimination (high \textsc{Resume}) yet are too conservative under genuine interruptions. Freeze-Omni offers the most balanced profile but over-reacts to background speech. Moshi shows instability across all non–floor-taking settings (\textsc{Unknown} high).

Timing-wise, effective human-like repair typically requires $t_{\text{resp}}\!\le\!1.5$\,s. GPT-4o achieves such gaps but often with the wrong action (false \textsc{Respond}), whereas Sonic and Gemini leave long gaps (2.0–2.7\,s) that may harm perceived conversational flow.

\noindent\textbf{Takeaways.} 
\textit{Best at interruptions:} GPT-4o (fast yielding), Freeze-Omni (shortest $t_{\text{resp}}$).  
\textit{Best at filtering:} Sonic (\textsc{backch}/\textsc{bkg\_speech}) and Gemini (\textsc{talking\_other}).  
Overall, systems optimized for \emph{fast yielding} must strengthen addressee detection and backchannel filtering, while those optimized for \emph{robust floor holding} must improve decisiveness under true interruptions.

\section{Prosodic and Quality Shifts}
% Beyond timing and categorical behaviors, we also examine prosodic adaptation—changes in rate, pitch, and intensity from pre- to post-overlap segments—to evaluate whether models produce natural and controlled re-entries.
Beyond categorical behaviors and timing, our benchmark also evaluates how models adapt their speech prosody and perceptual quality during overlap. Prosodic adaptation is a key marker of conversational competence: humans naturally modulate tempo, pitch, and energy when resuming after an interruption, signaling turn re-entry and preserving conversational naturalness.

Specifically, we investigate whether models modulate their delivery \textit{within} an utterance when responding to a user interruption.
For trials labeled \textsc{respond}, we compare the segment immediately before the overlap (\PreOverlap) with the repair segment after the overlap (\PostOverlap). We evaluate \emph{speaking rate} (WPM), \emph{pitch} (mean, standard deviation (SD)), \emph{intensity} (mean, standard deviation (SD)), and \emph{predicted MOS} (UTMOSv2) using paired $t$-tests ($p<0.05$). Results are summarized in Table~\ref{tab:intr_prepost}. To promptly respond to user interruptions, we expect slight increases in speaking rate (WPM) and a subtle pitch/energy lift at re-entry, without large intensity swings. MOS should not degrade; large intensity SD spikes or rate jumps indicate brittle control rather than skillful adaptation.

Comparing \PreOverlap{} to \PostOverlap{} on \textsc{respond} trials (Table~\ref{tab:intr_prepost}) reveals two dominant adaptation regimes. First, a \emph{tempo/pitch–lift} regime in which GPT-4o (and, to a lesser extent, Gemini) systematically accelerates and elevates pitch—accompanied for GPT-4o by greater pitch variability—consistent with an emphatic, floor-reacquisition strategy. Second, a \emph{soft-but-dynamic intensity} regime in which Freeze-Omni and Sonic reduce mean intensity while increasing intensity variability, yielding quieter yet more contoured restarts with minimal change in rate or pitch. Moshi departs from both patterns: it exhibits an extreme speed-up alongside decreases in pitch level and variability and a drop in mean intensity with higher intensity variability, suggesting hurried, less melodic, and brittle re-entry control. Predicted quality (UTMOSv2) is largely stable pre\,\(\to\)\,post, with only a small decline observed for Sonic.

\noindent\textbf{Takeaway.} Strong models make well-controlled increases in tempo and pitch to signal turn re-entry while keeping speech natural. In contrast, overly fast or uneven intensity changes lead to rushed and unstable responses. These results suggest that future systems should adopt prosodic adjustments to maintain fluency and conversational appropriateness.

\section{Conclusion}
\label{sec:conclusion}
Managing overlap is a core competency for real-time conversational AI. We present \textsc{Full-Duplex-Bench v1.5}, a fully automated, scenario-controlled benchmark that makes overlap measurable by formalizing expected behavior and timing (\(t_{\text{stop}}, t_{\text{resp}}\)) across four representative cases, enabling reproducible comparison beyond turn-based evaluation.
Across five state-of-the-art systems, we find a stable trade-off between \emph{responsiveness} and \emph{floor holding}: fast yielders excel on true interruptions but over-accommodate incidental speech, while robust holders resist non-addressed input yet delay necessary repairs. Our metrics provide scenario-specific latency targets and behavior distributions that make these differences comparable.
By open-sourcing tasks, metrics, and code, \textsc{Full-Duplex-Bench v1.5} offers a practical yardstick to diagnose weaknesses, track progress beyond half-duplex paradigms, and engineer systems that handle the fluid dynamics of human conversation with greater fluency and social awareness.

\section{Acknowledgments}
This work was supported by the Ministry of Education (MOE) of Taiwan under the project Taiwan Centers of Excellence in Artificial Intelligence, through the NTU Artificial Intelligence Center of Research Excellence (NTU AI-CoRE).

\bibliographystyle{IEEEbib}
\bibliography{refs}

\end{document}